\documentclass[fleqn,twoside]{article}
\usepackage[headings]{espcrc2}

\readRCS
$Id: espcrc2.tex,v 1.2 2004/02/24 11:22:11 spepping Exp $
\usepackage{graphicx}
\usepackage[figuresright]{rotating}
\usepackage{mathrsfs}

\newcommand{\AmS}{{\protect\the\textfont2
  A\kern-.1667em\lower.5ex\hbox{M}\kern-.125emS}}
\newcommand{\epem}{\ifmmode {e^+e^-} \else ${e^+e^-}$  \fi}
\newcommand{\ppbar}{\ifmmode {p \overline{p}} \else ${p \overline{p}}$  \fi}
\newcommand{\bbbar}{\ifmmode {b \overline{b}} \else ${b \overline{b}}$  \fi}

\newcommand{\dedx}{dE/dx}
\newcommand{\br}{\ensuremath{\mathcal{B}}}
\newcommand{\rfrac}{\frac{BR(B \rightarrow D^0_{flav} K)}{BR(B \rightarrow D^0_{flav} \pi)}}
\newcommand{\degs}{\ifmmode {^{\circ}} \else ${^{\circ}}$  \fi}
\newcommand{\micron}{\ifmmode {\mu m} \else ${\mu m}$  \fi}
\newcommand{\ipb}{\ifmmode {pb^{-1}} \else ${pb^{-1}}$  \fi}
\newcommand{\ifb}{\ifmmode {fb^{-1}} \else ${fb^{-1}}$  \fi}

\newcommand{\Bo}{B^{0}}

\newcommand{\bdkpi}{B^{0} \rightarrow K^{+} \pi^{-} }
\newcommand{\abdkpi}{\bar{B}^{0} \rightarrow K^{-} \pi^{+} }
\newcommand{\bdpipi}{B^{0} \rightarrow \pi^{+} \pi^{-} }
\newcommand{\bskpi}{B_s^{0} \rightarrow K^{-} \pi^{+} }
\newcommand{\bskk}{B_s^{0} \rightarrow K^{+} K^{-} }

\title{CP Violation Studies at Tevatron}

\author{E. Ben-Haim\address[MCSD]{LPNHE - Universite Pierre et Marie Curie-Paris6, UMR7585, Paris F-75005 France; IN2P3-CNRS}
        \thanks{On behalf of the CDF and D\O\ collaborations}}
       
\begin{document}

\begin{abstract}
We present an overview of a few recent results related to CP-violation from the Tevatron.
First, we discuss a measurement of the dimuon charge asymmetry from D\O\ , that extracts the
CP-violation parameter of $\Bo$ mixing and decay. This is followed by the CDF measurement
of the CP-violating asymmetry in $\bdkpi$ decays. Finally we give the CDF result on the ratio 
$R = \frac{BR(B \rightarrow D^0 K)}{BR(B \rightarrow D^0 \pi)}$.

\vspace{1pc}
\end{abstract}

\maketitle

\section{Introduction}

Heavy flavor physics at the Tevatron exploits the large $\bbbar$ production cross section and the fact that, unlike at the $B$-factories, all species of B hadrons are produced. There are also several challenges, mainly due to the huge total inelastic cross section and to the abundance of combinatorical background.
The analyses described below became possible thanks to the high performances of the CDF and D\O\ detectors.
CDF has an excellent mass resolution and the unique ability to trigger events
with charged particles  originated in vertices displaced from the
primary $\ppbar$ vertex (displaced tracks). The particle identification, with both the Time-of-Flight detector and $dE/dx$ from the drift chamber, is crucial.
D\O\ uses its superb muon system: large $(\eta, \phi)$ coverage, good scintillator based triggering and
cosmic ray rejection, low punch-through and precision tracking. The D\O\ analysis presented here uses the clean muon id and the ability to reverse the toroid and solenoid magnetic fields.

\section{CP-Violation Parameter of $\Bo$ Mixing and Decay}
The D\O\ experiment has extracted the CP-violation parameter
of $\Bo$ mixing and decay from the dimuon charge asymmetry.
This measurement used an integrated luminosity of $970$ pb$^{-1}$.
CP-violation in mixing, which has not yet been observed for $B$ mesons,
is sensitive to several extensions of the Standard Model \cite{randall,hewett}.
The CP-violation parameter $\epsilon_{B^0}$ can be obtained by measuring $A_{SL}$,
the asymmetry of the same side lepton pairs coming from direct $B$ 
decays\cite{randall,pdg}:
\begin{eqnarray}
\frac{4\Re{(\epsilon_{B^0})}}{1 + \left| \epsilon_{B^0} \right|^2} = A_{SL} = \nonumber \\
\frac{N(\bbbar \to l^{+}l^{+}X) - N(\bbbar \to l^{-}l^{-}X)}{N(\bbbar \to l^{+}l^{+}X) - N(\bbbar \to l^{-}l^{-}X)}.
\label{epsilon}
\end{eqnarray}
$A_{SL}$ is extracted from the dimuon charge asymmetry $A = \frac{N^{++} - N^{--}}{N^{++} + N^{--}}$,
where $N^{++}$ ($N^{--}$) is the number of events with two
positive (negative) muon candidates passing selection cuts.
The measured $A$ must contain only the physics part of the charged asymmetry,
separating it from detector effects. In order to relate $A$ to $A_{SL}$, 
all processes contributing to $A$ have to be identified.

The polarities of the toroid and solenoid magnetic fields are reversed
roughly every week so that the four solenoid-toroid polarity
combinations are exposed to approximately the same
integrated luminosity. This allows cancellation of first order
detector effects, as the possibly different reconstruction efficiencies
of positive and negative tracks due to different trajectories. 

\begin{figure}
\begin{center}
\vspace*{2.0cm}
\scalebox{0.4}
{\includegraphics[0in,1in][8in,9.5in]{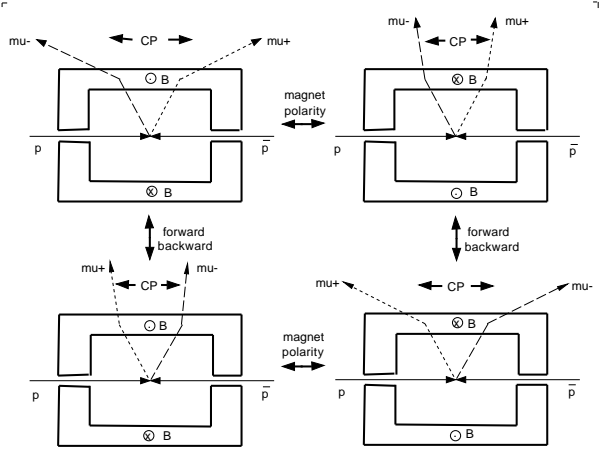}}
\vspace*{-6.0cm}
\caption{Schematic drawing of the magnetized iron toroids
of the D\O\ detector, and muon tracks related
by toroid polarity reversal, CP conjugation and
forward-backward reflection.
}
\vspace*{-1.0cm}
\label{detector}
\end{center}
\end{figure}

The muon detector is shown schematically in Figure \ref{detector}.
Let $n_\alpha^{\beta\gamma}$ be the number of muons passing cuts with charge
$\alpha = \pm 1$, toroid polarity $\beta = \pm 1$, and
$\gamma = +1$ if
$\eta > 0$ and $\gamma = -1$ if $\eta < 0$. 
The physics and
the detector are modeled as follows:
\begin{eqnarray}
n_\alpha^{\beta\gamma} \equiv \frac{1}{4}
N \epsilon^\beta (1 + \alpha A)(1 + \alpha \gamma A_{fb})
(1 + \gamma A_{det}) \nonumber \\
\mbox{} \times (1 + \alpha\beta\gamma A_{ro}) 
(1 + \beta \gamma A_{\beta \gamma})
(1 + \alpha \beta A_{\alpha \beta}).
\label{asymmetries}
\end{eqnarray}
$N$ is the number of muons passing cuts, and
$\epsilon^\beta$ is the fraction of integrated luminosity with
toroid polarity $\beta$ ($\epsilon^+ + \epsilon^- = 1$).
Equation (\ref{asymmetries}) defines six asymmetries.
$A$ is the dimuon charge asymmetry, $A_{fb}$ is the forward-backward
asymmetry (that quantifies the tendency of $\mu^+$ to go in
the proton direction and $\mu^-$ to go in the anti-proton
direction), $A_{det}$ 
measures the north-south asymmetry
of the detector, and $A_{ro}$ is the range-out asymmetry
(that quantifies the change in acceptance and range-out of
muon tracks that bend toward, or away from, the iron toroid magnet).
$A_{\alpha \beta}$ is a detector asymmetry between
tracks bending north and tracks bending south.
$A$ and $A_{fb}$ are physics
asymmetries that we want to measure, and $A_{det}$,
$A_{ro}$ and $A_{\alpha \beta}$ are detector asymmetries.
The model (\ref{asymmetries}) fits 8 numbers
$n_\alpha^{\beta \gamma}$ with 8 parameters ($N$, $\epsilon^+$,
and 6 asymmetries).

The measured value of $A$ is
\begin{equation}
A = -0.0013 \pm 0.0012\textrm{(stat)} \pm 0.0008\textrm{(sys)}.
\label{final_A_5}
\end{equation}
The resulting CP-violation parameter, taking into account the weights of the different physics processes producing charged dimuon pairs, and assuming that the asymmetry (if any) is due to
asymmetric $B^0 \leftrightarrow \bar{B}^0$ mixing and decay is
\begin{eqnarray*}
\frac{\Re{(\epsilon_{B^0})}}{1 + \left| \epsilon_{B^0} \right|^2}=
-0.0011 \pm 0.0010\textrm{(stat)} \pm 0.0007\textrm{(sys)}.
\label{CPV_singlemu_1}
\end{eqnarray*}
This is the world best measurement of $\epsilon_{B^0}$. The dominating systematic error is due to prompt $\mu$ + $K^\pm$-decay.

\section{CP asymmetry in $\bdkpi$ decays}

The CDF experiment has measured the CP-violating asymmetry in $\bdkpi$ decays using an integrated luminosity of $360$ pb$^{-1}$ approximately.
The flavor-specific $\bdkpi$ decay 
occurs in the SM through the dominant tree and penguin diagrams.  Their interfering amplitudes induce the
CP asymmetry $A_{CP}(\bdkpi)$ defined as follows:
\begin{equation}
\frac{\br(\abdkpi) - \br(\bdkpi)}{\br(\abdkpi) + \br(\bdkpi)} 
\end{equation}
$B$-factories recently measured a $\mathcal{O}(10\%)$ asymmetry with 2\% accuracy, probing for the first time direct CP violation 
in the $b$-quark sector \cite{acp_1,acp_2}; however, additional experimental information is needed because theoretical predictions still 
suffer from large (5--10\%) uncertainties \cite{acp_3,acp_4,acp_5}, and the observed asymmetries in neutral and charged modes are not consistent, 
as the SM would suggest. A measurement from the Tevatron is therefore interesting, also for the unique possibility to combine asymmetry measurements
in $\bdkpi$ and $\bskpi$ decays, which provide a model-independent probe for the presence of non-SM physics \cite{acp_6}.

\begin{figure}[ht]
\centering
\includegraphics[width=80mm]{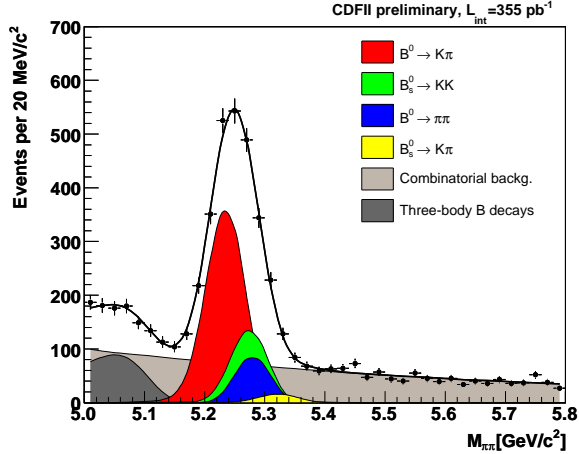}
\vspace*{-1.5cm}
\caption{Invariant $\pi\pi$-mass
after the optimized offline selection with individual signal components (cumulative) and backgrounds (overlapping) overlaid. Contributions to background come from random pairs of tracks which satisfy the selection requirements and partially-reconstructed $B^0_{(s)}$ decays, as resulting from the invariant-mass fit.}
\vspace*{-0.8cm}
 \label{fig:acp_1}
\end{figure}

The sample of pairs of oppositely-charged particles reconstructed with $\pi$ mass assignment, has been used to form $B^0_{(s)}$ meson candidates. After applying a set of optimized selection cuts, the resulting $\pi\pi$-mass distribution (Histogram in Figure \ref{fig:acp_1}) shows a clean signal. 

Despite the excellent mass resolution, the various $B^0_{(s)}\to h^{+}h'{-}$ modes overlapped into an unresolved mass peak, while the PID resolution was
insufficient for separating them on an event-by-event basis. We achieved a statistical separation instead, with a
 multivariate, unbinned likelihood-fit (fit of composition) that used PID information, provided by the $\dedx$ in the drift chamber, and kinematics. The fit used five observables: 
the invariant $\pi\pi$-mass $m_{\pi\pi}$, the signed momentum-imbalance $\alpha=(1-p_1/p_2)q_1$, the scalar sum of particles' momenta $p_{tot}$, and the $\dedx$ of both particles. In $\alpha$, the momentum (charge) $p_1$ ($q_1$) refers to the softer track.
By combining kinematics and charge information, the fit separated also 
$K^+\pi^-$ from  $K^-\pi^+$ final states.

The fit found three modes contributing to the peak:
$313 \pm 34$ $\bdpipi$, $1475 \pm 60$ $\bdkpi$,  and $523 \pm 41$ $\bskk$ decays. A not yet statistically significant contribution
of $64 \pm 30$ $\bskpi$ decays was also found. Fit projections are overlaid to data in Figure \ref{fig:acp_1}.
From $787 \pm 42$ reconstructed $\bdkpi$ decays and $689 \pm 41$ reconstructed $\abdkpi$ decays, we measured the following uncorrected
 value for the direct CP asymmetry:
\begin{eqnarray}
\frac{N(\abdkpi) - N(\bdkpi)}{N(\abdkpi) + N(\bdkpi)} =  \nonumber \\
(-6.6 \pm 3.9)\%. 
\end{eqnarray}
Above result was then corrected for differences in trigger, reconstruction, and selection efficiencies between $\bdkpi$ and $\abdkpi$ modes

The dominant source of the systematic error was the uncertainty  on the $\dedx$ model for kaons, pions, and track-to-track correlation.  This effect is expected to partially reduce as the size of 
of $D$ meson decays samples, used for calibration of the $\dedx$ model, increases. The second important contribution derived from the statistical uncertainty on the nominal value of $B^0_{s}$ masses.
Since we use the $B^0_{s}$ masses measured by CDF,  this uncertainty will reduce with the increasing statistic of fully-reconstructed $B^0_{s}$ decays.

We quote the following result for the direct CP asymmetry in $\bdkpi$ decays, where all contributions to 
the systematic uncertainty have been summed in quadrature:
\begin{equation}
A_{CP}(\bdkpi) = (-5.8 \pm 3.9 \pm 0.7)\%,
\end{equation} 
which is approximately $1.5\sigma$ different from zero, and in agreement with world best results: 
$A_{CP}(\bdkpi) = (-10.8 \pm 2.4 \pm 0.7)\%$=
from the Babar Collaboration \cite{acp_1} and
$A_{CP}(\bdkpi) = (-9.3 \pm 1.8 \pm 0.8)\%$,
from the Belle Collaboration \cite{acp_2}.

CDF result is still limited by the statistic uncertainty; however, its systematic uncertainty, at the same level of
$B$-factories, is  promising:  with significantly more data already collected,  we expect to reduce the statistical uncertainty 
 down to 2.5\%, which will make CDF result competitive with $B$-factories soon.

\section{Measurement of the ratio $\frac{BR(B \rightarrow D^0 K)}{BR(B \rightarrow D^0 \pi)}$}

This CDF measurement exploited an integrated luminosity of $360$ pb$^{-1}$ approximately to measure the ratio $R = \rfrac$. $R$ is a necessary input of the GLW method \cite{glw1,glw2} to obtain the CKM angle $\gamma$.

\begin{figure}[htb] 
  \centering
  \includegraphics[width=36mm]{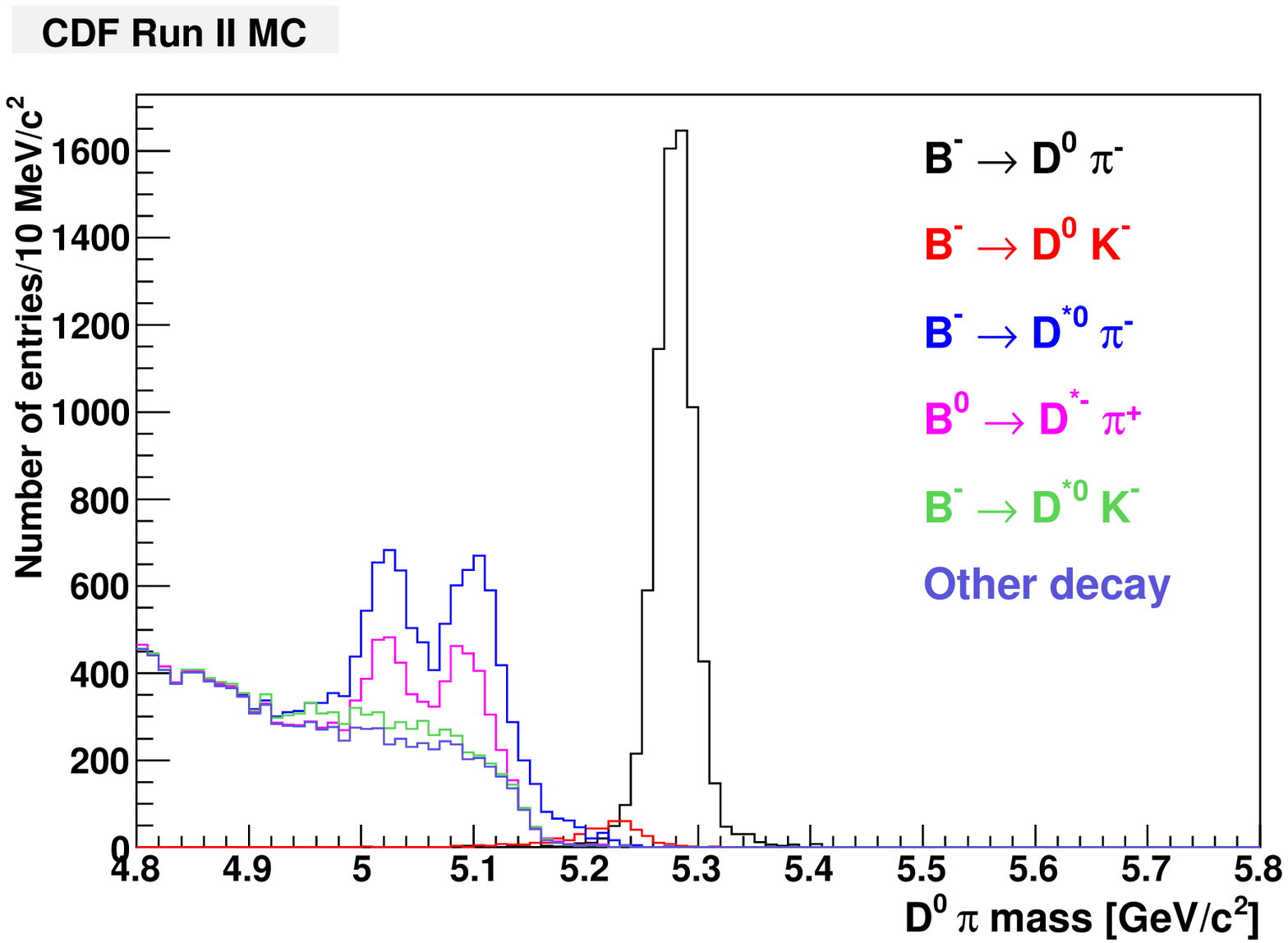} 
  \includegraphics[width=36mm]{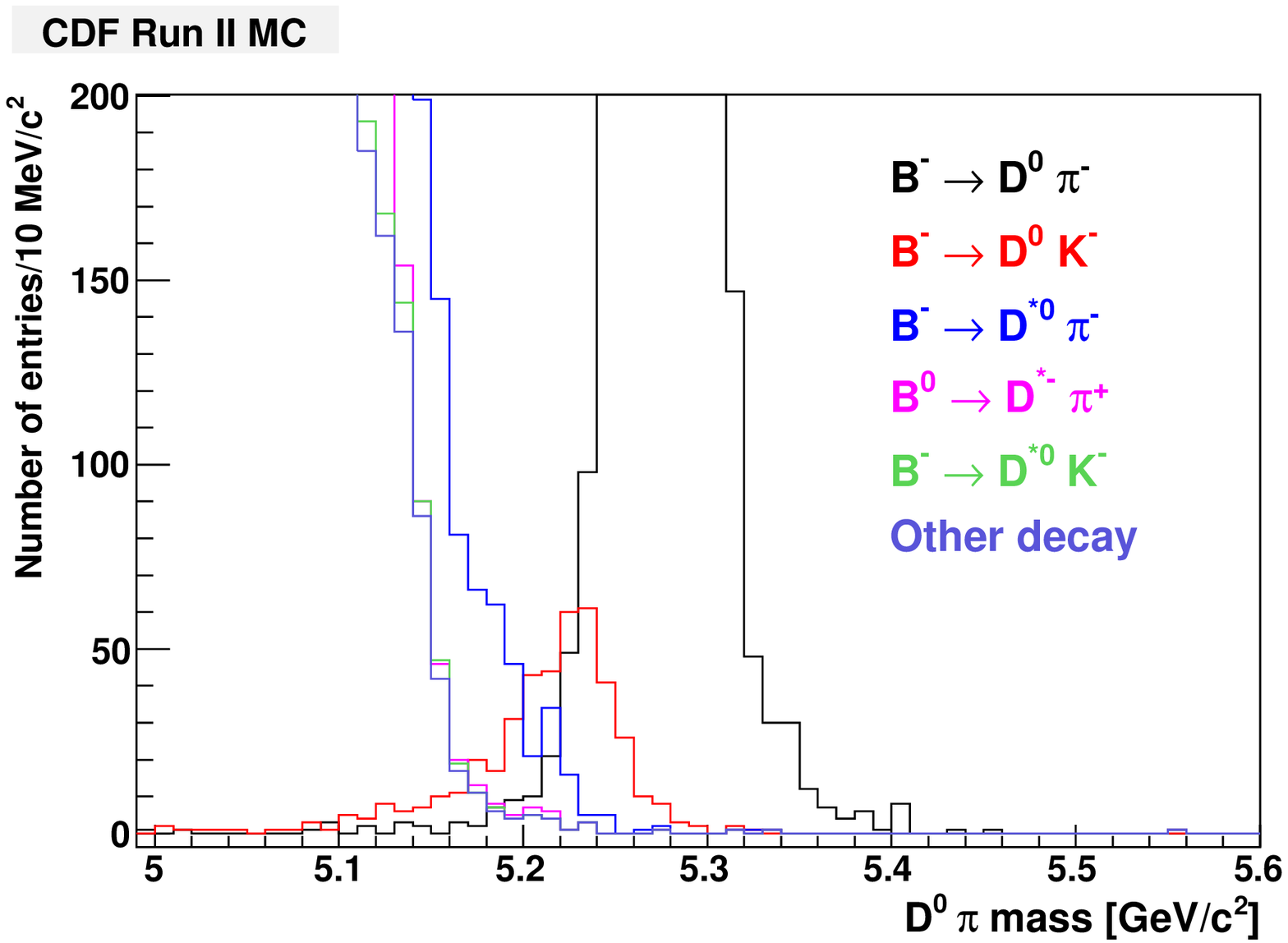}
  \vspace*{-0.8cm}
\caption{Invariant $D\pi$ mass obtained from generic $B^{+}$ MC sample, showing the different signal and background reconstructed components:  $B^+ \rightarrow \overline{D}^0 \pi^+$, $B^+ \rightarrow \overline{D}^0 K^+$, $B^+ \rightarrow \overline{D}^{0*} \pi^+$, $B^+ \rightarrow D^{*-} \pi^+$, $B^+ \rightarrow \overline{D}^{0*} K^+$ and other decay modes. Right: zoom on the suppressed $B^+ \rightarrow \overline{D}^0 K^+$ peak region.}
\vspace*{-0.5cm}
  \label{MC}
\end{figure}

\begin{figure}[htb] 
  \centering
  \includegraphics[width=80mm]{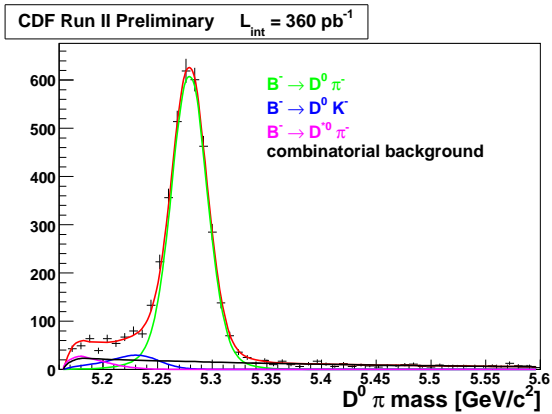} 
  \vspace*{-1.4cm}
  \caption{Mass Projections of the multivariate likelihood fit}
  \vspace*{-0.5cm}
  \label{proj_m}
\end{figure}

We reconstructed the $B^- \rightarrow D^0 \pi^-$ with $D^0 \rightarrow K^- \pi^+$ (flavour eigenstate) and the $B^- \rightarrow D^0_{CP+} \pi^-$ with $D^0_{CP+} \rightarrow \pi^+ \pi^-$ and $D^0_{CP+} \rightarrow K^+ K^-$ (CP-even eigenstate). The $\pi$ mass is assigned to the $B$ daughter track. Selection cuts have been applied. 
The resulting invariant $D^0\pi$ mass from a generic Monte Carlo sample is shown in Figure \ref{MC}, where various $B \to D^{(*)}h^{\pm}$ are present. To reject most of the background contributions while keeping the $B^+ \rightarrow \overline{D}^0 K^+$ signal, we used the narrow mass window $5.17<m(D\pi)<5.6$. The resulting $D^0\pi$ mass in data is the histogram in Figure \ref{proj_m}.

We separated the yields of $B^- \rightarrow {D^0} K^-$ from $B^- \rightarrow 
{D^0} \pi^-$ and the different background contributions using an multivariate, unbinned likelihood-fit, exploiting information provided by the $dE/dx$ in the drift chamber and kinematics. Fit projections are overlaid to data in Figure \ref{proj_m}.

The raw fit results ($3265 \pm 38$ $B \to D\pi$ and $224 \pm 22$ $B \to DK$ decays) have been corrected for detector effects and analysis efficiencies in order to obtain the final result. We quote:
\begin{eqnarray}\label{ratio}
  R = \frac{BR(B\rightarrow D^0_{flav} K)}{BR(B\rightarrow D^0_{flav} \pi)} = \nonumber \\
   0.065 \pm 0.007(stat) \pm 0.004(sys).
\end{eqnarray}
The world average for $R$ is $0.0830 \pm 0.0035$. It combines the results from the Belle ($0.077 \pm 0.005 \pm 0.006$ \cite{belle}), Babar ($0.0831 \pm 0.0035 \pm 0.002$ \cite{babar}) and Cleo ($0.099^{+0.014 + 0.007}_{-0.012 -0.006}$ \cite{cleo}) collaborations.

\end{document}